\documentclass[10pt,journal]{IEEEtran}
\usepackage{graphicx, color, psfrag}
\usepackage{amsmath, amssymb, amstext}
\usepackage{fancyhdr}
\usepackage{subeqnarray}
\usepackage{cite}
\usepackage{hhline}
\usepackage{slashbox}
\usepackage{multirow}

\begin{document}

%%%%%%%%%%%%%%%%%%%%%%%%%%%%%%%%%%%%%%%%%%%%%%%%%%%%%%%%%%%%%%%%%%%%%%%%%%%%%%%%%%%%%%%%%%%%%%%%%%%%%%%%%%%%%%%%%%%%%%%%%%%%%%%%%%%%%%%
% Beginning of title
\title{High SNR BER Comparison of Coherent and Differentially Coherent Modulation Schemes in Lognormal Fading Channels}

\author{Xuegui Song,~\IEEEmembership{Student~Member,~IEEE}, Julian Cheng,~\IEEEmembership{Senior~Member,~IEEE}, \\and Mohamed-Slim Alouini,~\IEEEmembership{Fellow,~IEEE}
\thanks{X.~Song and J.~Cheng are with School of Engineering, The University of British Columbia, Kelowna, BC, Canada (e-mail: \{xuegui.song, julian.cheng\}@ubc.ca).}% <-this % stops a space
\thanks{M.-S. Alouini is with the Electrical Engineering Program, Computer, Electrical and Mathematical Sciences and Engineering (CEMSE) Division, King Abdullah University of Science and Technology (KAUST), Thuwal, Makkah Province, Saudi Arabia (e-mail: slim.alouini@kaust.edu.sa).}
}

\maketitle

%%%%%%%%%%%%%%%%%%%%%%%%%%%%%%%%%%%%%%%%%%%%%%%%%%%%%%%%%%%%%%%%%%%%%%%%%%%%%%%%%%%%%%%%%%%%%%%%%%%%%%%%%%%%%%%%%%%%%%%%%%%%%%%%%%%%%%%
% Begin of abstract
\begin{abstract}
Using an auxiliary random variable technique, we prove that binary differential phase-shift keying and binary phase-shift keying have the same asymptotic bit-error rate performance in lognormal fading channels. We also show that differential quaternary phase-shift keying is exactly $2.32$ dB worse than quaternary phase-shift keying over the lognormal fading channels in high signal-to-noise ratio regimes.
\end{abstract}
% End of abstract
%%%%%%%%%%%%%%%%%%%%%%%%%%%%%%%%%%%%%%%%%%%%%%%%%%%%%%%%%%%%%%%%%%%%%%%%%%%%%%%%%%%%%%%%%%%%%%%%%%%%%%%%%%%%%%%%%%%%%%%%%%%%%%%%%%%%%%%


%%%%%%%%%%%%%%%%%%%%%%%%%%%%%%%%%%%%%%%%%%%%%%%%%%%%%%%%%%%%%%%%%%%%%%%%%%%%%%%%%%%%%%%%%%%%%%%%%%%%%%%%%%%%%%%%%%%%%%%%%%%%%%%%%%%%%%%
% Begin of Introduction
\section{Introduction}
\label{sec:introduction}
Digital communication systems in wireless channels have been studied extensively (refer to \cite{SimonBook}, \cite{ProakisBook}, and references therein). Besides the well-investigated Rayleigh, Rician, and Nakagami-$m$ channels, the lognormal fading is widely recognized as an important channel model because it fits empirical fading measurements well in many transmission scenarios of practical interest \cite{Lotse1992}. For examples, the lognormal fading can characterize the shadowing effects in outdoor RF communications \cite{SimonBook} and describe indoor radio propagation environments. The lognormal fading is also suitable for describing ultra-wideband channels \cite{Molisch2003} and characterizing optical wireless communication (OWC) links in clear sky over several hundred meters \cite{EJLee2004}.
%The lognormal distribution can also describe slowly-varying channel gains in the mixed small- and large-scale effects exhibiting dominant lognormal statistics \cite{Alouini2002}. In addition, the lognormal fading is suitable for describing ultra-wideband channels \cite{Molisch2003} and characterizing optical wireless communication (OWC) links in clear sky over several hundred meters \cite{EJLee2004}.

On the other hand, performance analysis of digital communication systems over the lognormal channels remains challenging and mathematically intractable. The error rate analysis of such systems typically involves numerical integration \cite{Zhu2002} or approximation \cite{Liu2003}. In \cite{R.1}, the authors studied the achievable finite diversity order of the lognormal channels in relatively low signal-to-noise ratio (SNR) regimes. However, their result is not valid when the SNR is asymptotically large since the diversity order for the lognormal channels is widely known to be $\infty$. Due to this fact, the asymptotic theory \cite{ZhengdaoWang2003} cannot be applied directly to the lognormal channels.

Binary phase-shift keying (BPSK) is an important benchmark modulation scheme whose bit-error rate (BER) performance over fading channels is well-known \cite{SimonBook}, \cite{ProakisBook}. For coherent demodulation, BPSK requires channel state information (CSI) at the receiver in order to achieve
its optimal performance. When the required CSI cannot be tracked accurately, binary differential
phase-shift keying (DPSK) is an attractive alternative with inferior BER performance. It is
well-known that DPSK is 3 dB worse than BPSK in Rayleigh channels in high SNR regimes \cite{Ekanayake1990}.
One fundamental question we aim to answer is what the performance loss of DPSK with respect to (w.r.t.) BPSK will be in the lognormal channel at high SNR.

In this letter, we prove that in the lognormal fading DPSK and BPSK have the same asymptotic BER performance, and differential quaternary phase-shift keying (DQPSK) is exactly $2.32$ dB worse than quaternary phase-shift keying (QPSK) in high SNR regimes.
%%%%%%%%%%%%%%%%%%%%%%%%%%%%%%%%%%%%%%%%%%%%%%%%%%%%%%%%%%%%%%%%%%%%%%%%%%%%%%%%%%%%%%%%%%%%%%%%%%%%%%%%%%%%%%%%%%%%%%%%%%%%%%%%%%%%%%%
% End of Introduction

\section{General Asymptotic Relative BER Analysis}
\label{sec-analysis}

For a digital communication system over fading channels, the SNR can be expressed as $\gamma = \overline{\gamma}I^2$ where $\overline{\gamma}$ is the average SNR\footnote{In RF communications, $\overline{\gamma}$ is the average SNR. In OWC with direct detection, $\overline{\gamma}$ is known as the electrical SNR \cite{Zhu2002}.} and $I$ is the channel gain. The asymptotic BER $P_b^{\infty}$ as $\overline{\gamma}\to\infty$ can be obtained by averaging the conditional BER $P_b(\gamma)$ over the probability density function (pdf) of $\gamma$, and it is approximated by \cite {ZhengdaoWang2003}
\begin{equation}
f(\gamma) = \frac {c \gamma^{t}}{\overline{\gamma}^{t+1}} + o\left({\gamma^{t}}\right)
\label{equ-GeneralPDFofSNR}
\end{equation}
where a function $f(x)$ is $o(g(x))$ if $\lim\limits_{x\to 0} f(x)/g(x) = 0$. Equation \eqref{equ-GeneralPDFofSNR} can be obtained from the Taylor series expansion of the SNR pdf where the parameters $c$ and $t$ are finite constants. For BPSK with conditional BER $P_{b,\mbox{\tiny BPSK}}(\gamma) = Q\left(\sqrt{2{\gamma}} \right)$ where $Q(x)=\int_{x}^\infty e^{-z^2/2}/\sqrt{2\pi}dz$ is the Gaussian $Q$-function, the asymptotic BER is \cite {ZhengdaoWang2003}
\begin{equation}
\label{equ-SERforCoherent}
P_{b,\mbox{\tiny BPSK}}^{\infty} = \frac{c \Gamma\left(t+\frac{3}{2}\right)}{2\sqrt{\pi}\left(t+1\right){\overline{\gamma}}^{t+1}} + o\left(\frac {1} {\overline{\gamma}^{t+1}}\right)
\end{equation}
where $\Gamma(\cdot)$ is the Gamma function. For DPSK with conditional BER $P_{b,\mbox{\tiny DPSK}}(\gamma) = 0.5\exp(-{\gamma})$, the asymptotic BER is \cite{Li2012}
\begin{equation}
\label{equ-SERforNonCoherent}
P_{b,\mbox{\tiny DPSK}}^{\infty} = \frac{c \Gamma\left(t+2\right)} {2(t+1){\overline{\gamma}}^{t+1}}
+ o\left(\frac {1} {\overline{\gamma}^{t+1}}\right).
\end{equation}
Using \eqref{equ-SERforCoherent} and \eqref{equ-SERforNonCoherent}, we derive the asymptotic SNR penalty factor of DPSK w.r.t. BPSK as
\begin{equation}
\label{SNR-BPSKDPSK}
{\mbox{SNR}}_{\mbox{\tiny DPSK-BPSK}}^{\infty} = \frac{10}{t+1}\log\left(\frac{\sqrt{\pi}\Gamma(t+2)}{\Gamma\left(t+\frac{3}{2}\right)}\right) \mbox{dB}
\end{equation}
where $\log(\cdot)$ is the log function with base $10$.

Since QPSK and BPSK have the same BER performance \cite{ProakisBook}, we have $P_{b,\mbox{\tiny QPSK}}^{\infty} = P_{b,\mbox{\tiny BPSK}}^{\infty}$. For Gray coded DQPSK, the conditional BER is given as \cite[Eq. (8.86)]{SimonBook}
\begin{equation}
\label{P_4DPSK}
P_{b,\mbox{\tiny{DQPSK}}}(\gamma) = F\left(\frac{\pi}{4}, \gamma\right) -F\left(\frac{5\pi}{4}, \gamma\right)
\end{equation}
where
\begin{equation}
\label{F_function}
F(\psi, \gamma) = \frac{\sin\psi}{2\pi}\int_{0}^{\pi/2}\frac{\exp\left(-2\gamma(1-\cos\psi\cos\theta)\right)}{1-\cos\psi\cos\theta} d\theta.
\end{equation}
Using \eqref{equ-GeneralPDFofSNR}, \eqref{P_4DPSK}, and \eqref{F_function}, we obtain the asymptotic BER as
\begin{equation}
\label{P_4DPSK_asy}
P_{b,\mbox{\tiny{DQPSK}}}^{\infty} = \frac{c\Gamma\left(t+2\right)\left[g\left(t,\frac{\pi}{4}\right)+g\left(t,\frac{5\pi}{4}\right)\right]}{2^{t+2}\sqrt{2}\pi(t+1)\overline{\gamma}^{t+1}} + o\left(\frac {1} {\overline{\gamma}^{t+1}}\right)
\end{equation}
where $g(t,\psi) \triangleq \int_0^{\pi/2}\frac{1}{(1-\cos\psi\cos\theta)^{t+2}}d\theta$. Using \eqref{equ-SERforCoherent} and \eqref{P_4DPSK_asy}, we derive the asymptotic SNR penalty factor of DQPSK w.r.t. QPSK as
\begin{equation}
\label{SNR-QPSKDQPSK}
{\mbox{SNR}}_{\mbox{\tiny DQPSK-QPSK}}^{\infty}\hspace{-0.06cm} = \hspace{-0.06cm}\frac{10}{t+1}\hspace{-0.06cm}\log\hspace{-0.06cm}\left(\hspace{-0.04cm}\frac{\Gamma(t+2)\left[g\left(t,\frac{\pi}{4}\right)+g\left(t,\frac{5\pi}{4}\right)\right]}{2^{t+1}\sqrt{2\pi}\Gamma\left(t+\frac{3}{2}\right)}\hspace{-0.04cm}\right) \mbox{dB}.
\end{equation}
Equations \eqref{SNR-BPSKDPSK} and \eqref{SNR-QPSKDQPSK} are general results for BPSK/DPSK and QPSK/DQPSK over fading channels when the receiver SNR pdf can be expressed as \eqref{equ-GeneralPDFofSNR}. From both equations, we observe that the SNR penalty factors are only functions of $t$.

%%%%%%%%%%%%%%%%%%%%%%%%%%%%%%%%%%%%%%%%%%%%%%%%%%%%%%%%%%%%%%%%%%%%%%%%%%%%%%%%%%%%%%%%%%%%%%%%%%%%%%%%%%%%%%%%%%%%%%%%%%%%%%%%%%%%%%%
% End of Turbulence Model

%%%%%%%%%%%%%%%%%%%%%%%%%%%%%%%%%%%%%%%%%%%%%%%%%%%%%%%%%%%%%%%%%%%%%%%%%%%%%%%%%%%%%%%%%%%%%%%%%%%%%%%%%%%%%%%%%%%%%%%%%%%%%%%%%%%%%%%
% Begin of Analysis
\section{Asymptotic Relative BER Analysis for the Lognormal Fading Channels}
\label{sec:performanceanalysis}

%%%%%%%%%%%%%%%%%%%%%%%%%%%%%%%%%%%%%%%%%%%%%%%%%%%%%%%%%%%%%%%%%%%%%%%%%%%%%%%%%%%%%%%%%%%%%%%%%%%%%%%%%%%%%%%%%%%%%%%%%%%%%%%%%%%%%%%
\subsection{Single-Input Single-Output System}
\label{subsec:SISO}
In a lognormal fading environment, the channel gain $I = \exp(X)$ where $X$ is a Gaussian random variable (RV) with mean $\mu$ and variance $\sigma^2$. Thus, $I$ follows a lognormal distribution. To facilitate our
analysis, we normalize the mean of $I$ (i.e., $E[I] = \exp(\mu + \sigma^2/2) = 1$ and thus $\mu = -\sigma^2/2$) and obtain the pdf of $I$ as
\begin{equation}
\label{lognormalPDF}
f_{LN}(I) = \frac{1}{\sqrt{2\pi}\sigma I}\exp\left(-\frac{(\ln I+\sigma^2/2)^2}{2\sigma^2}\right).
\end{equation}
The parameter $\sigma$ is associated with the severity of fading. For OWC applications, the typical value of $\sigma$ is less than $0.5$ and it can be as low as $0.02$ \cite{Zhu2002}.
%One can show that the amount of fading (AF) is $\mbox{AF}=\exp(4\sigma^2)-1$.

%using $P_b = \int_0^\infty P_b(\gamma)f(\gamma)d\gamma$ where $P_b(\gamma)$ is the conditional BER
In order to evaluate the average BER performance, we require pdf of the receiver SNR. This pdf can be obtained by changing variables in \eqref{lognormalPDF} as
\begin{align}
\label{SNRPDF}
f_{LN}(\gamma)
&= \frac{1}{2\sqrt{2\pi}\sigma \gamma}\exp\left(-\frac{(\ln \gamma+\sigma^2-\ln\overline{\gamma})^2}{8\sigma^2}\right)
\end{align}
which is another lognormal pdf. Our goal is to study the asymptotic relative BER performance between coherent and differentially coherent modulation schemes over the lognormal channels. While it is infeasible to perform asymptotic analysis directly on the lognormal channels, we introduce the lognormal-Nakagami fading as an auxiliary channel model where the receiver SNR follows a lognormal-Gamma distribution. Unlike the lognormal pdf, the lognormal-Gamma pdf has a Taylor series expansion at the origin so that one can perform asymptotic analysis on such a channel. It can be shown that the diversity order of the lognormal-Nakagami channel is the parameter $m$. As $m$ approaches $\infty$, the lognormal-Gamma pdf approaches that of a lognormal RV. Therefore, we can study the asymptotic relative BER performance between coherent and differentially coherent modulation schemes over the lognormal-Nakagami channels and obtain the limiting results for the lognormal channels.

For the lognormal-Nakagami channels, the receiver SNR $\gamma$ follows a lognormal-Gamma distribution with pdf
\begin{equation}
\begin{split}
\label{SNRPDFGamma}
f_{LG}(\gamma) =& \int_0^{\infty}\frac{m^m\gamma^{m-1}}{\Omega^m\Gamma(m)}\exp\left(-\frac{m\gamma}{\Omega}\right)  \\
&\hspace{-0.8cm}\times\frac{1}{2\sqrt{2\pi}\sigma \Omega}\exp\left(-\frac{(\ln \Omega+\sigma^2-\ln\overline{\gamma})^2}{8\sigma^2}\right)d\Omega
\end{split}
\end{equation}
where $m$ is the Nakagami-$m$ parameter and $\Omega$ is the second moment of a Nakagami-$m$ RV. When $m\to\infty$, we have
\begin{equation}
\label{limit}
\lim_{m\to\infty}\frac{m^m\gamma^{m-1}}{\Omega^m\Gamma(m)}\exp\left(-\frac{m\gamma}{\Omega}\right) = \delta\left(\frac{\gamma}{\Omega}-1\right)
\end{equation}
where $\delta(\cdot)$ is the Dirac delta function. Applying \eqref{limit} to \eqref{SNRPDFGamma}, we obtain
\begin{equation}
\lim_{m\to\infty}f_{LG}(\gamma) = \frac{1}{2\sqrt{2\pi}\sigma \gamma}\exp\left(-\frac{(\ln \gamma+\sigma^2-\ln\overline{\gamma})^2}{8\sigma^2}\right)
\end{equation}
which, as expected, is the lognormal pdf given in \eqref{SNRPDF}.

When $\gamma\to 0^+$, we can show that
\begin{equation}
\begin{split}
\label{SNRPDFGamma_re}
f_{LG}(\gamma) =& \frac{m^m{\gamma}^{m-1}}{2\sqrt{2\pi}\sigma\Gamma(m)}\int_0^{\infty}\frac{1}{\Omega^{m+1}}  \\
&\hspace{-0.9cm}\times\exp\left(-\frac{(\ln \Omega+\sigma^2-\ln \overline{\gamma})^2}{8\sigma^2}\right)d\Omega + o\left({\gamma^{m-1}}\right).
\end{split}
\end{equation}
Comparing \eqref{SNRPDFGamma_re} with \eqref{equ-GeneralPDFofSNR}, we obtain $t=m-1$ so that the diversity order of the lognormal-Nakagami channels is $m$. Substituting $t=m-1$ into \eqref{SNR-BPSKDPSK}, we obtain the asymptotic relative BER performance between DPSK and BPSK over the lognormal-Nakagami channels as
\begin{equation}
\label{SNR_LG}
{\mbox{SNR}}_{\mbox{\tiny DPSK-BPSK}}^{\infty}(m) = \frac{10}{m}\log\left(\frac{\sqrt{\pi}\Gamma(m+1)}{\Gamma\left(m+\frac{1}{2}\right)}\right) \mbox{dB}.
\end{equation}
When $m\to\infty$, it follows that\footnote{Mathematically, the $0$ dB asymptotic BER performance loss of DPSK w.r.t. BPSK in \eqref{SNR_LG_infty} can also be obtained by letting $t\to\infty$ in \eqref{SNR-BPSKDPSK}. However, such manipulation is not permissible because the asymptotic analysis theory requires $t$ be finite.}
\begin{equation}
\label{SNR_LG_infty}
\lim_{m\to\infty}\hspace{-0.03cm}{\mbox{SNR}}_{\mbox{\tiny DPSK-BPSK}}^{\infty}(m)\hspace{-0.05cm} =\hspace{-0.09cm} \lim_{m\to\infty}\hspace{-0.04cm}\frac{10}{m}\log\hspace{-0.04cm}\left(\hspace{-0.03cm}\frac{\sqrt{\pi}\Gamma(m+1)}{\Gamma\left(m+\frac{1}{2}\right)}\hspace{-0.03cm}\right)\hspace{-0.04cm}=0\ \mbox{dB}
\end{equation}
which indicates that the BER of DPSK over the lognormal fading channels will approach that of BPSK in high SNR regimes. In another word, there is no substantial benefit to choose BPSK over DPSK when operating on the lognormal fading channels in asymptotically high SNR regimes.

Substituting $t=m-1$ into \eqref{SNR-QPSKDQPSK}, we obtain the asymptotic BER performance loss of DQPSK w.r.t. QPSK over the lognormal-Nakagami channels as
\begin{equation}
\label{SNR_QPSK_LG}
\begin{split}
{\mbox{SNR}}_{\mbox{\tiny DQPSK-QPSK}}^{\infty}(m) &=\left[ \frac{10}{m}\log\left(\frac{\Gamma(m+1)}{\sqrt{2\pi}\Gamma\left(m+\frac{1}{2}\right)}\right) -10\log(2)\right. \\
&\hspace{-1.2cm} \left.+\frac{10}{m}\log\left(g\left(m-1,\frac{\pi}{4}\right)+g\left(m-1,\frac{5\pi}{4}\right)\right)\right]\;\mbox{dB}.
\end{split}
\end{equation}
From the definition of $g(t,\psi)$, we have
\begin{equation}
g\left(m-1,{5\pi}/{4}\right)=\int_0^{\pi/2}{\left(1+{\sqrt{2}}\cos\theta/{2}\right)^{-(m+1)}}d\theta.
\end{equation}
It can be shown that $\lim\limits_{m\to\infty}g\left(m-1,{5\pi}/{4}\right) = 0$ when $\theta\in[0, \pi/2]$ so that the term $g\left(m-1,{5\pi}/{4}\right)$ in \eqref{SNR_QPSK_LG} is negligible when $m\to\infty$. Defining $h(\theta)\triangleq{1}/({1-{\sqrt{2}}\cos\theta/{2}})$, we have
\begin{equation}
\begin{split}
&\lim_{m\to\infty}\frac{10}{m}\log\left(g\left(m-1,\frac{\pi}{4}\right)\right) \\
&\hspace{0.8cm}= {10}\log\left(\lim_{m\to\infty}\left[\int_0^{\pi/2}{\left(h(\theta)\right)^{m+1}}d\theta\right]^{\frac{1}{m}}\right) \\
&\hspace{0.8cm}= {10}\log\left(\lim_{m\to\infty}\left[\int_0^{\pi/2}{\left(h(\theta)\right)^{m}}d\theta\right]^{\frac{1}{m}}\right) \\
&\hspace{0.8cm}=10\log\left(\left\| h \right\|_{\infty}\right)
\end{split}
\end{equation}
where $\left\| h \right\|_{\infty}$ is the uniform norm of the function $h(\theta),\,\theta\in[0, \pi/2]$. Since $h(\theta)$ is a continuous function on $[0, \pi/2]$, we have $\left\| h \right\|_{\infty}= \max \{h(\theta): \theta\in[0, \pi/2]\} = 2+\sqrt{2}$. It follows that the asymptotic BER performance loss of DQPSK w.r.t. QPSK over the lognormal channels is
\begin{equation}
\label{SNR_QPSK_LN}
\begin{split}
\lim_{m\to\infty}{\mbox{SNR}}_{\mbox{\tiny DQPSK-QPSK}}^{\infty}(m)  &= 10\log\left(2+\sqrt{2}\right) - 10\log(2)  \\
                                                                  &= 2.32\; \mbox{dB}.
\end{split}
\end{equation}

%%%%%%%%%%%%%%%%%%%%%%%%%%%%%%%%%%%%%%%%%%%%%%%%%%%%%%%%%%%%%%%%%%%%%%%%%%%%%%%%%%%%%%%%%%%%%%%%%%%%%%%%%%%%%%%%%%%%%%%%%%%%%%%%%%%%%%%
\subsection{Selection Combining Diversity System}
\label{subsec:SC}
The asymptotic relative performance predictions in Section \ref{subsec:SISO} may not occur in practical SNR range.
However, diversity technique can facilitate the convergence of the relative BER performance in practical SNR range
to the asymptotic predictions. We will show such effect of diversity technique by studying an $L$-branch selection
combining (SC) system over independent and identically distributed (i.i.d.) lognormal channels. The pdf of
the SC output SNR $\gamma_{sc}$ is
\begin{equation}
\label{SNRPDF_SC}
\begin{split}
f_{LN}(\gamma_{sc}) =& Lf_{LN}(\gamma_{sc})\left[F_{LN}(\gamma_{sc})\right]^{L-1}                   
\end{split}
\end{equation} 
where $F_{LN}(\gamma)$ is the cumulative distribution function of $\gamma$, and it can be obtained from \eqref{SNRPDF} as
\begin{equation}
\label{SNRCDF}
F_{LN}(\gamma) = 1-Q\left(\frac{\ln \gamma+\sigma^2-\ln\overline{\gamma}}{2\sigma}\right).
\end{equation}
Therefore, one can obtain the average BER of such a system by substituting \eqref{SNRPDF_SC}
and \eqref{SNRCDF} into $P_b = \int_0^\infty P_b(\gamma_{sc})f(\gamma_{sc})d\gamma_{sc}$.

Since the diversity order of an $L$-branch SC system is $L\varphi$ assuming $\varphi$ is the diversity order of a single-input single-output system \cite{Li2012}, the diversity parameter of such an $L$-branch system over the lognormal-Nakagami channels is $t=mL-1$. Substituting $t=mL-1$ into \eqref{SNR-BPSKDPSK} and letting $m\to\infty$, we obtain the asymptotic BER performance loss of DPSK w.r.t. BPSK over the lognormal channels as $0$ dB. Similarly, we obtain the asymptotic BER performance loss of a DQPSK system w.r.t. a QPSK system over the lognormal channels as $2.32$ dB by substituting $t=mL-1$ into \eqref{SNR-QPSKDQPSK} and letting $m\to\infty$.

%%%%%%%%%%%%%%%%%%%%%%%%%%%%%%%%%%%%%%%%%%%%%%%%%%%%%%%%%%%%%%%%%%%%%%%%%%%%%%%%%%%%%%%%%%%%%%%%%%%%%%%%%%%%%%%%%%%%%%%%%%%%%%%%%%%%%%%
\subsection{Multiple-Input Multiple-Output System}

The analysis in Section \ref{subsec:SC} can be extended to an $M\times N$ multiple-input multiple-output (MIMO) system because the maximum achievable diversity order of such a system is $MN\varphi$ \cite{Zheng2003TCOM}. It is straightforward to obtain that the diversity parameter $t=mMN-1$ for an $M\times N$ system over the lognormal-Nakagami channels. Therefore, using \eqref{SNR-BPSKDPSK} we obtain the asymptotic relative BER performance between an MIMO system using BPSK and that using DPSK as $0$ dB over the lognormal channels. The relative performance is still $2.32$ dB between an MIMO system using QPSK and that using DQPSK over the lognormal channels. However, MIMO technique will facilitate the convergence of the relative BER performance in practical SNR range to the asymptotic predictions.

%%%%%%%%%%%%%%%%%%%%%%%%%%%%%%%%%%%%%%%%%%%%%%%%%%%%%%%%%%%%%%%%%%%%%%%%%%%%%%%%%%%%%%%%%%%%%%%%%%%%%%%%%%%%%%%%%%%%%%%%%%%%%%%%%%%%%%%
% End of Analysis
%%%%%%%%%%%%%%%%%%%%%%%%%%%%%%%%%%%%%%%%%%%%%%%%%%%%%%%%%%%%%%%%%%%%%%%%%%%%%%%%%%%%%%%%%%%%%%%%%%%%%%%%%%%%%%%%%%%%%%%%%%%%%%%%%%%%%%%

\begin{figure}
\begin{center}
\includegraphics[width=0.95\linewidth, draft=false]{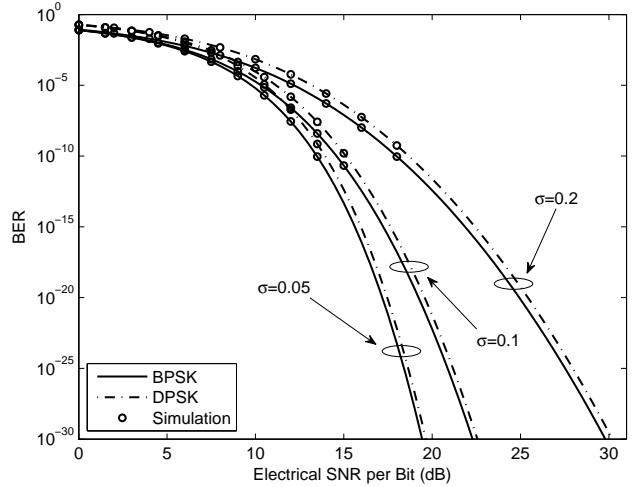}
\caption{BERs of subcarrier intensity modulated OWC systems using BPSK and DPSK over the lognormal fading channels.}
\label{Fig_BERLognormal}
\end{center}
\end{figure}

\begin{figure}
\begin{center}
\includegraphics[width=0.95\linewidth, draft=false]{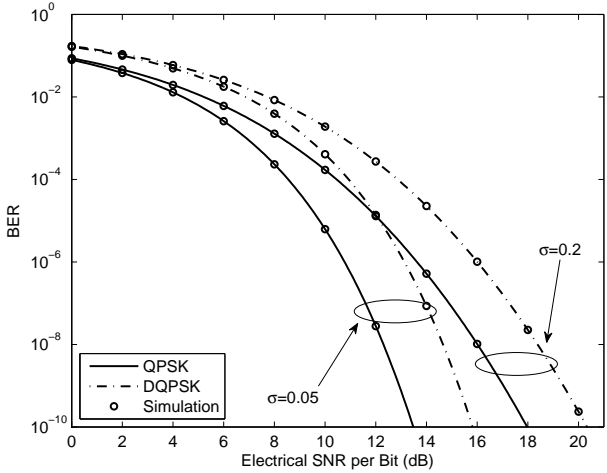}
\caption{BERs of subcarrier intensity modulated OWC systems using QPSK and DQPSK over the lognormal fading channels.}
\label{Fig_BERQPSK_LN}
\end{center}
\end{figure}

%%%%%%%%%%%%%%%%%%%%%%%%%%%%%%%%%%%%%%%%%%%%%%%%%%%%%%%%%%%%%%%%%%%%%%%%%%%%%%%%%%%%%%%%%%%%%%%%%%%%%%%%%%%%%%%%%%%%%%%%%%%%%%%%%%%%%%%
% End of Figures

%%%%%%%%%%%%%%%%%%%%%%%%%%%%%%%%%%%%%%%%%%%%%%%%%%%%%%%%%%%%%%%%%%%%%%%%%%%%%%%%%%%%%%%%%%%%%%%%%%%%%%%%%%%%%%%%%%%%%%%%%%%%%%%%%%%%%%%
% Begin of Numerical Results
\section{Numerical Results}
\label{sec:numerical-results}

In Fig. \ref{Fig_BERLognormal}, we present BER curves of subcarrier OWC systems using BPSK and DPSK
over the lognormal channels with different $\sigma$ values. A close examination of Fig. \ref{Fig_BERLognormal}
indicates that the BER performance loss of subcarrier DPSK w.r.t. subcarrier BPSK is $0.7$ dB
at the BER level of $10^{-10}$ when $\sigma=0.2$. When $\sigma=0.05$, this performance loss is
reduced to $0.2$ dB at the BER level of $10^{-30}$. When the SNR value is asymptotically large,
this performance loss will vanish as predicted by \eqref{SNR_LG_infty}. In Fig. \ref{Fig_BERLognormal} the exact BER curves are obtained via numerical integration, and they are validated by Monte Carlo simulations.

From the numerical results presented in Fig. \ref{Fig_BERLognormal}, we observe that the $0$ dB asymptotic prediction will not occur in practical SNR regimes. In order to investigate the performance loss in practical SNR range, through numerical investigation, we present in Table \ref{table1} the observed BER performance loss values at different BER levels in the lognormal channels. From Table \ref{table1}, we observe that the BER performance loss depends on the parameter $\sigma$ and it decreases in weaker lognormal channels at a given BER level. We also observe a clear trend that the performance loss decreases when the SNR increases and it will approach the $0$ dB prediction when the SNR value is asymptotically large. For example, the performance loss is as low as $0.63$ dB at SNR level around $14$ dB when $\sigma=0.1$, and it further decreases to $0.48$ dB at SNR level around $13$ dB when $\sigma=0.05$. We conclude that this BER performance loss is small in practical SNR range although it cannot achieve the $0$ dB asymptotic
prediction.

Figure \ref{Fig_BERQPSK_LN} presents BER curves of subcarrier OWC systems using QPSK and DQPSK over the lognormal channels with different $\sigma$ values. A close examination of Fig. \ref{Fig_BERQPSK_LN} indicates that the BER performance loss of subcarrier DQPSK w.r.t. subcarrier QPSK is $2.4$ dB at the BER level of $10^{-10}$ when $\sigma=0.2$ and this performance loss reduces to $2.3$ dB when $\sigma=0.05$. These numerical results agree with the asymptotic prediction of $2.32$ dB from \eqref{SNR_QPSK_LN}.

In order to demonstrate the effect of diversity technique on the relative performance in practical SNR range, through numerical investigation, we present in Table \ref{table2} the required SNR range to achieve a fixed performance gap of $0.5$ dB for different $L$ values in i.i.d. lognormal channels. From Table \ref{table2}, we observe that in order to achieve a fixed performance gap of $0.5$ dB for a given $\sigma$, the required SNR range depends on the number of branch $L$, and these SNR values will decrease when $L$ increases. When $\sigma =0.2$ and $L=4$, the $0.5$ dB performance gap can be achieved in a practical SNR range of $13.6-14.1$ dB. A clear trend can be observed from Table \ref{table2} that diversity technique can facilitate the convergence of the BERs of BPSK and DPSK at high SNR. When the SNR value is sufficiently large, this performance loss is expected to vanish for an $L$-branch SC system.

%%%%%%%%%%%%%%%%%%%%%%%%%%%%%%%%%  Beginning of Table%%%%%%%%%%%%%%%%%%%%%%%%%%%%%%%%
\begin{table}
\renewcommand{\arraystretch}{1.3}
\caption{Performance loss of DPSK with respect to BPSK in {\normalfont dB} at different BER levels in the lognormal channels}% with different $\sigma$ values}
\label{table1}
\centering
\begin{tabular}{|c|c|c|c|c|c|c|}
\hline
\hline
\backslashbox{$\sigma$}{BER}  & $10^{-2}$   & $10^{-4}$  & $10^{-6}$  & $10^{-8}$ & $10^{-10}$ \\
\hline
\hline
 $0.5$           &$2.29$   & $1.65$   & $1.38 $  & $1.24$   & $1.10$    \\
 $0.2$           &$1.84$   & $1.56$   & $0.89 $  & $0.79$   & $0.68$    \\
 $0.1$           &$1.67$   & $0.98$   & $0.75 $  & $0.63$   & $0.54$    \\
 $0.05$          &$1.61$   & $0.92 $  & $0.66 $  & $0.55$   & $0.48$    \\
\hline
\hline
%\hline
\end{tabular}
\end{table}

%%%%%%%%%%%%%%%%%%%%%%%%%%%%%%%%%%%%%%%%%%%%%%%%%%%%%%%%%%%%%%%%%%%%%%%%%%%%%%%%%%%%%%%%%%%%%%%%%%%%%%%%%%%%%%%%%%%%%%%%%%%%%%%%%%%%%%%
% End of Numerical Results

%%%%%%%%%%%%%%%%%%%%%%%%%%%%%%%%  Beginning of Table%%%%%%%%%%%%%%%%%%%%%%%%%%%%%%%%
\begin{table}
\renewcommand{\arraystretch}{1.3}
\caption{The required SNR ({\normalfont dB}) range to achieve a fixed performance gap of $0.5$ {\normalfont dB} for different $L$ values}
\label{table2}
\centering
\begin{tabular}{|c|c|c|c|c|c|}
\hline
\hline
\backslashbox{$\sigma$}{$L$}  & $1$   & $3$  & $4$  & $5$  \\
\hline
\hline
 $0.2$           &$27.5-28.0$   & $16.8-17.3$   & $13.6-14.1 $  & $12.8-13.3$    \\
 $0.1$           &$14.6-15.1$   & $11.9-12.4$   & $11.7-12.2 $  & $11.5-12.0$      \\
 %$0.05$          &$1.61$   & $0.92 $  & $0.66 $  & $0.55$      \\
\hline
\hline
\end{tabular}
\end{table}

%%%%%%%%%%%%%%%%%%%%%%%%%%%%%%%%%%%%%%%%%%%%%%%%%%%%%%%%%%%%%%%%%%%%

%%%%%%%%%%%%%%%%%%%%%%%%%%%%%%%%%%%%%%%%%%%%%%%%%%%%%%%%%%%%%%%%%%%%%%%%%%%%%%%%%%%%%%%%%%%%%%%%%%%%%%%%%%%%%%%%%%%%%%%%%%%%%%%%%%%%%%%
\section{Conclusions}
\label{sect:conclusion}
Asymptotic analyses show that the performance advantage of BPSK over DPSK in the lognormal channels vanishes in asymptotically large SNR regimes. Under this operating condition, DPSK is a preferred modulation scheme in weak lognormal channels because it does not require CSI estimation and tracking. The performance advantage of QPSK over DQPSK in the lognormal channels is quantified to be $2.32$ dB in large SNR regimes. We also demonstrated that diversity technique can facilitate the convergence of the relative BER performance in practical SNR range to the asymptotic predictions.

%%%%%%%%%%%%%%%%%%%%%%%%%%%%%%%%%%%%%%%%%%%%%%%%%%%%%%%%%%%%%%%%%%%%%%%%%%%%%%%%%%%%%%%%%%%%%%%%%%%%%%%%%%%%%%%%%%%%%%%%%%%%%%%%%%%%%%%
% Begin of Bibliography
\bibliographystyle{IEEEtran}

%%%%%%%%%%%%%%%%%%%%%%%%%%%%%%%%%%%%%%%%%%%%%%%%%%%%%%%%%%%%%%%%%%%%%%%%%%%%%%%%%%%%%%%%%%%%%%%%%%%%%%%%%%%%%%%%%%%%%%%%%%%%%%%%%%%%%%%
% End of Bibliography

\end{document}